\documentstyle[epsfig,amsmath,12pt]{article}

\author{Petr Z\'avada\\
\\\
{\it Institute of Physics, Academy of Sciences of Czech Republic}\\
{\it Na Slovance 2, CZ-180 40 Prague 8}\\
E-mail: zavada@fzu.cz}
\title{Nucleon spin structure and mass of quarks
}
\date{May 31, 1999
}
\begin{document}

\maketitle
\begin{abstract}
The alternative to the standard formulation of the quark-parton model is
proposed. Our relativistically covariant approach is based on the solution
of the master equations relating the structure and distribution functions,
which consistently takes into account the intrinsic quark motion - in
contradistinction to the standard infinite momentum approach, in which this
motion is latently suppressed. The model well reproduces the experimental
data on the both polarized and unpolarized structure functions, assuming
that only the valence quarks term contributes to the nucleon spin. It is
shown, the combined analysis of the polarized and unpolarized data can give
an information about the effective masses and intrinsic motion of the quarks
inside the nucleon. Simultaneously, it is shown that the rate of the nucleon
energy carried by the quarks can be less, than estimated from the standard
approach. As an addition, a prediction for the proton spin function $g_2$ is
given.
\end{abstract}

\renewcommand{\theequation} {\thesection .\arabic{equation}}

\section{Introduction}

\label{sec1}

Measuring of the structure functions is an unique tool for the study of the
nucleon internal structure - together with the quark-parton model (QPM)
giving the relations between the structure functions and the parton
distributions, which represent a final, detailed picture of the nucleon. In
this sense, these relations, obtained under definite assumptions, are
extraordinary important, since the distribution functions themselves are not
directly measurable. At the same time, the standard, simple formulas
relating the structure and distribution functions are ordinarily considered
so self-evident, that in some statements, the both are identified.

The experiments dedicated to the Deep Inelastic Scattering (DIS), are
oriented to the measuring either {\it unpolarized} or {\it polarized}
structure functions. The results on the unpolarized functions are 
compatible with our expectations based on the QPM and QCD, but the situation
for the polarized functions is much more complicated. Until now, it is not
well understood, why the integral of the proton spin structure function $g_1$
is substantially less, than expected from very natural but possibly equally
naive assumption, that the nucleon spin is created by the valence quarks.
Presently, there is a strong tendency to explain the missing part of the
nucleon spin as a result of the considerable contributions of the sea quarks
(particularly strange quarks) and the gluons. Nevertheless, a consistent
explanation of the underlying mechanism is still missing. During the last
years, the hundreds of papers have been devoted to the nucleon spin
structure, for the present status see e.g. \cite{jer}, \cite{dis98}, the
comprehensive overviews \cite{ans},\cite{hai} and citations therein.

In the present paper we continue our discussion started in \cite{zav1}, \cite
{zav2}, where we have shown, that the standard formulation of the QPM,
conceptually firmly connected with the infinite momentum frame (IMF),
oversimplifies the parton kinematics. In \cite{zav2} we demonstrated that
the effect of oversimplified kinematics in IMF can have an impact
particularly on the spin structure function $g_1$, or more exactly, it can
substantially modify the relation between the distribution and spin
structure functions. We have suggested this effect can be a source of the
discrepancy between the experimentally measured and naturally expected
magnitude of the spin function $g_1$. The primary aim of this paper is to
precise just this point.

The paper is organized as follows. In the following section we add some
comments to our master equations connecting the structure and distribution
functions and point out the distinction between them and these used in the
standard approach. In the Sec. \ref{sec3}. we propose the model, in which
the internal motion of quarks is consistently taken into account. Further,
in contradistinction to the standard treatment based on the QCD evolution of
the distribution functions in dependence on $Q^2$, our model rather aims at
describing the part of distribution and structure functions, which are not
calculable in terms of the pertubative QCD. In the Sec. \ref{sec4}. the
results of the model on the polarized and unpolarized proton structure
functions are compared with the experimental data and some free parameters
are fixed. Some important additional comments on the model and obtained
results are done in the Sec. \ref{sec5}. The last section is devoted to the
overall summary and concluding remarks. Since this paper should be read
together with \cite{zav1}, \cite{zav2}, for convenience we refer to the
equations or figures in these papers simply with prefixes P,Q.

For completeness, in the Appendix we give a more rigorous explanation, that
the approximation of the term 
\begin{equation}
\label{app0}\frac{pq}{M\nu }\simeq \frac{p_0+p_1}M 
\end{equation}
entering the argument of the $\delta -$functions in some integrals in this
and two the previous papers is for our purpose fully acceptable.

\section{Master equations}

\label{sec2}

\setcounter{equation}{0}

In the preceding discussion \cite{zav1},\cite{zav2} we have shown [see Eqs.
(P3.41), (Q2.1)], that if one assumes momenta distributions (Q2.2), (Q2.3)
of quasi-free quarks having mass $m$ are spherically symmetric in the
nucleon rest frame, then the corresponding structure functions $%
W_1,W_2,G_1,G_2$obey the master equation%
$$
P_\alpha P_\beta \frac{W_2}{M^2}-g_{\alpha \beta }W_1+i\epsilon _{\alpha
\beta \lambda \sigma }q^\lambda \left[ s^\sigma MG_1+(Pqs^\sigma -sqP^\sigma
)\frac{G_2}M\right] +A(P_\alpha q_\beta +P_\beta q_\alpha )+Bq_\alpha
q_\beta 
$$
$$
=\int G(p_0)(2p_\alpha p_\beta -g_{\alpha \beta }\,pq)\delta ((p+q)^2-m^2) 
\frac{d^3p}{p_0} 
$$
\begin{equation}
\label{ms21}+i\epsilon _{\alpha \beta \lambda \sigma }q^\lambda \int
H(p_0)mw^\sigma \delta ((p+q)^2-m^2)\frac{d^3p}{p_0}. 
\end{equation}
Simultaneously, we have shown, how the explicit solution of this equation
looks like. In this solution, the mass $x_0=m/M$ formally appears as a free
parameter. Let us make a remark, in the same way a similar equation can be
obtained and solved also for the set of the neutrino structure functions,
nevertheless in this paper we shall deal only with the electromagnetic ones.

Let us note, despite of the fact that Eq. (\ref{ms21}) is assembled for
quark momenta distributions $G,H$ in the nucleon rest frame, the equation is
relativistically covariant. Its manifestly covariant form follows
immediately from (\ref{ms21}) after the substitution 
\begin{equation}
\label{ms22}G(p_{0,lab})=G\left( \frac{pP}M\right) ,\qquad
H(p_{0,lab})=H\left( \frac{pP}M\right) . 
\end{equation}
For moving nucleon we have 
\begin{equation}
\label{ms23}P=(P_0,P_1,0,0),\qquad \beta =\frac{P_1}{P_0},\qquad \gamma = 
\frac{P_0}M 
\end{equation}
and 
\begin{equation}
\label{ms24}\frac{pP}M=\frac{p_0P_0-p_1P_1}M=\gamma (p_0-\beta
p_1)=p_{0,lab} 
\end{equation}
which means, that the phase space of the subset of quarks with $p_{0,lab}$
fixed, represented by the sphere 
\begin{equation}
\label{ms25}p_1^2+p_2^2+p_3^2=p_{0,lab}^2-m^2 
\end{equation}
in the nucleon rest frame, is in a boosted system correctly represented by
the ellipsoid with the shape defined by the Lorentz transform (\ref{ms24}).

Further, the Eq. (\ref{ms21}) involve also the quark polarization vector $%
w_\mu $, which as follows from \cite{zav2}, plays the crucial role for the
evaluation of the spin structure functions $g_1,g_2.$ So here we consider
desirable to derive its form in a more rigorous way, than we suggested for
Eq. (Q2.12). Generally, quark polarization vector should be constructed from
the proton momentum $P$, proton polarization vector $s$ and the quark
momentum $p$: 
\begin{equation}
\label{cr1}w_\mu =AP_\mu +Bs_\mu +Cp_\mu , 
\end{equation}
where $A,B,C$ are invariant functions of $P,s,p.$ Then contracting of Eq.
(Q2.5) with $P_\sigma ,q_\sigma $ and $s_\sigma $ and using relations (Q2.4)
give the equations 
\begin{equation}
\label{cr2}-sqMG_2=\frac m{2M\nu }\int H\left( \frac{pP}M\right) \left(
AM^2+CpP\right) \delta \left( \frac{pq}{M\nu }-x\right) \frac{d^3p}{p_0}, 
\end{equation}
\begin{equation}
\label{cr3}sqMG_1=\frac m{2M\nu }\int H\left( \frac{pP}M\right) \left( AM\nu
+Bsq+Cpq\right) \delta \left( \frac{pq}{M\nu }-x\right) \frac{d^3p}{p_0}, 
\end{equation}
\begin{equation}
\label{cr4}-MG_1-\nu G_2=\frac m{2M\nu }\int H\left( \frac{pP}M\right)
\left( -B+Cps\right) \delta \left( \frac{pq}{M\nu }-x\right) \frac{d^3p}{p_0}%
. 
\end{equation}
Elimination of $G_1,G_2$ gives 
\begin{equation}
\label{cr5}\int H\left( \frac{pP}M\right) Cp\left( \frac{\nu P/M-q}{sq}%
-s\right) \delta \left( \frac{pq}{M\nu }-x\right) \frac{d^3p}{p_0}=0 
\end{equation}
and since $P,q,s$ are independent, $C$ must be zero. The remaining
invariants $A,B$ follow from Eq. (Q2.4), which implies 
\begin{equation}
\label{cr6}A^2M^2-B^2=-1,\qquad ApP+Bps=0 
\end{equation}
and solution of these equations reads 
\begin{equation}
\label{cr7}A=\mp \frac{ps}{\sqrt{(pP)^2-(ps)^2M^2}},\qquad B=\pm \frac{pP}{
\sqrt{(pP)^2-(ps)^2M^2}}. 
\end{equation}
So the quark polarization vector has the form 
\begin{equation}
\label{cr8}w_\mu =\pm \frac{(pP)s_\mu -(ps)P_\mu }{\sqrt{(pP)^2-(ps)^2M^2}}. 
\end{equation}
Contributions of both possible solutions [sign $+(-)$ means that quark spin
is parallel (antiparallel) to the proton spin in its rest frame] are in our
calculation taken into account by the difference in Eq. (Q2.3). Apparently,
for the proton rest frame and polarization $s=(0,1,0,0)$ the last equation
is identical to Eq. (Q2.12). Further, obtained invariants $A,B,C$ now give
spin structure functions from Eqs. (\ref{cr2}), (\ref{cr3}) in covariant
form 
\begin{equation}
\label{cr9}G_1=\frac m{2M(Pq)(sq)}\int H\left( \frac{pP}M\right) \frac{%
(pP)(sq)-(ps)(Pq)}{\sqrt{(pP)^2-(ps)^2M^2}}\delta \left( \frac{pq}{Pq}%
-x\right) \frac{d^3p}{p_0}, 
\end{equation}
\begin{equation}
\label{cr10}G_2=\frac{mM}{2(Pq)(sq)}\int H\left( \frac{pP}M\right) \frac{ps}{
\sqrt{(pP)^2-(ps)^2M^2}}\delta \left( \frac{pq}{Pq}-x\right) \frac{d^3p}{p_0}%
. 
\end{equation}
Apparently, according to these relations the structure functions can depend
also on mutual orientation of $s$ and $q.$ Of course, this dependence is
more complicated, apart the factor $sq$ ahead of the integrals, integration
involves the terms $ps$ and $pq$. This question is being studied and will be
discussed in a separate paper. In the present paper our considerations will
be based on the results obtained in \cite{zav2}, which follow from Eqs. (\ref
{cr9}), (\ref{cr10}) applied in the proton rest frame for the longitudinal
polarization $s=(0,1,0,0)$. Obviously, for this case the last two equations
are equivalent to Eqs. (Q2.13), (Q2.14).

The scheme based on the Eqs. (\ref{ms21}) and (\ref{cr8}) with all their
implications suggested in \cite{zav1},\cite{zav2} can be a priori valid for
quasi-free quarks (on mass shell) filling up the nucleon volume. In this
sense the scheme represents a covariant formulation of the naive QPM. We
have shown that Eq. (\ref{ms21}) in which the quark internal motion is
consistently taken into account imply relations between the structure and
distribution functions different from those obtained in the standard
procedure relying on the IMF, which is based on the approximation $p_\mu
=x_\mu P$ (parton momenta components are $x$ fractions of the proton momenta
components). In fact this approximation in the covariant formulation is
equivalent to the assumption, that the partons are static with respect to
the nucleon, therefore there are suppressed not only the transversal
momenta, but also longitudinal ones. Of course, this consequence is somewhat
obscured just in the IMF, where all the relative motion is frozen, since all
the processes run infinitely slowly - including the passing of the probing
lepton through the nucleon. Let us remark, the standard relations (e.g. $%
F_2=x\sum e_i^2q_i$) obtained in the naive QPM with static quarks are
currently applied even in the standard approach based on QCD improved QPM,
which is not a consistent procedure, since it means that correct dynamics is
combined with incorrect kinematics.

In this way we have shown, that the relations between the structure and
distribution functions can be, at least on the level of the naive QPM,
strongly modified (particularly for the polarized case) by the parton
internal motion. This result can be instructive by itself. The impact of the
quark intrinsic motion on the function $g_1(x)$ has been discussed also in
some other approaches \cite{bo1}-\cite{rit} and necessity of the covariant
formulation for the spin structure functions has been pointed out in \cite
{jack}.

However, the real nucleon is much more complex object, than just a bag of
quasi-free fermions. But we shall try to assume the following. The relations
obtained within the scheme suggested above can be used as a good
approximation even for the interacting quarks, but provided that the term
'mass of quasi-free parton' is substituted by the term 'parton effective
mass'. By this term we mean the mass, which a free parton would have to have
to interact with the probing photon equally as the real, bounded one.
Intuitively, this mass should correlate to $Q^2$: a lower $Q^2$ allows more
time and space for the struck parton to interact with some others, as a
result the energy is transferred to a greater system than the parton itself.
And on contrary, the higher $Q^2$ should mediate interaction with more
''isolated'' parton. Moreover, we accept that the value of the effective
mass even for a fixed $Q^2$ can fluctuate - e.g. in a dependence on the
actual QCD process accompanying the photon momentum transfer. This means,
that the terms in the relations involving the mass of quasi-free parton $%
x_0=m/M$ will be substituted by their convolution with some 'mass
distribution' $\mu :$%
\begin{equation}
\label{cr11}f(x_0)\rightarrow \int \mu (x_0,Q^2)f(x_0)dx_0. 
\end{equation}

In the following we shall propose a simple, but sufficiently general model
for the unknown distributions $\mu ,G,H,$ in which all the dynamics of the
system is absorbed. Then, these distributions will be used as an input for
the calculating of the corresponding structure functions.

\section{Model}

\label{sec3}

\setcounter{equation}{0}

Construction of the model is based on the following assumptions and
considerations:

1) Parton distribution $P(\epsilon )d\epsilon $ representing the number of
quarks in the energy interval $\left\langle \epsilon ,\epsilon +d\epsilon
\right\rangle $ can be formally expressed : 
\begin{equation}
\label{mss31}P(\epsilon )=\sum_jr_jj\rho _j(\epsilon ),\qquad \sum_jr_j=1, 
\end{equation}
where $r_j$ is a probability that the nucleon is in the state with $j$
charged partons (quarks + antiquarks) of various flavors, and $\rho _j$ is
the corresponding average one-parton distribution, which fulfills 
\begin{equation}
\label{mss32}\int \rho _j(\epsilon )d\epsilon =1. 
\end{equation}

\begin{figure}
\begin{center}
\epsfig{file=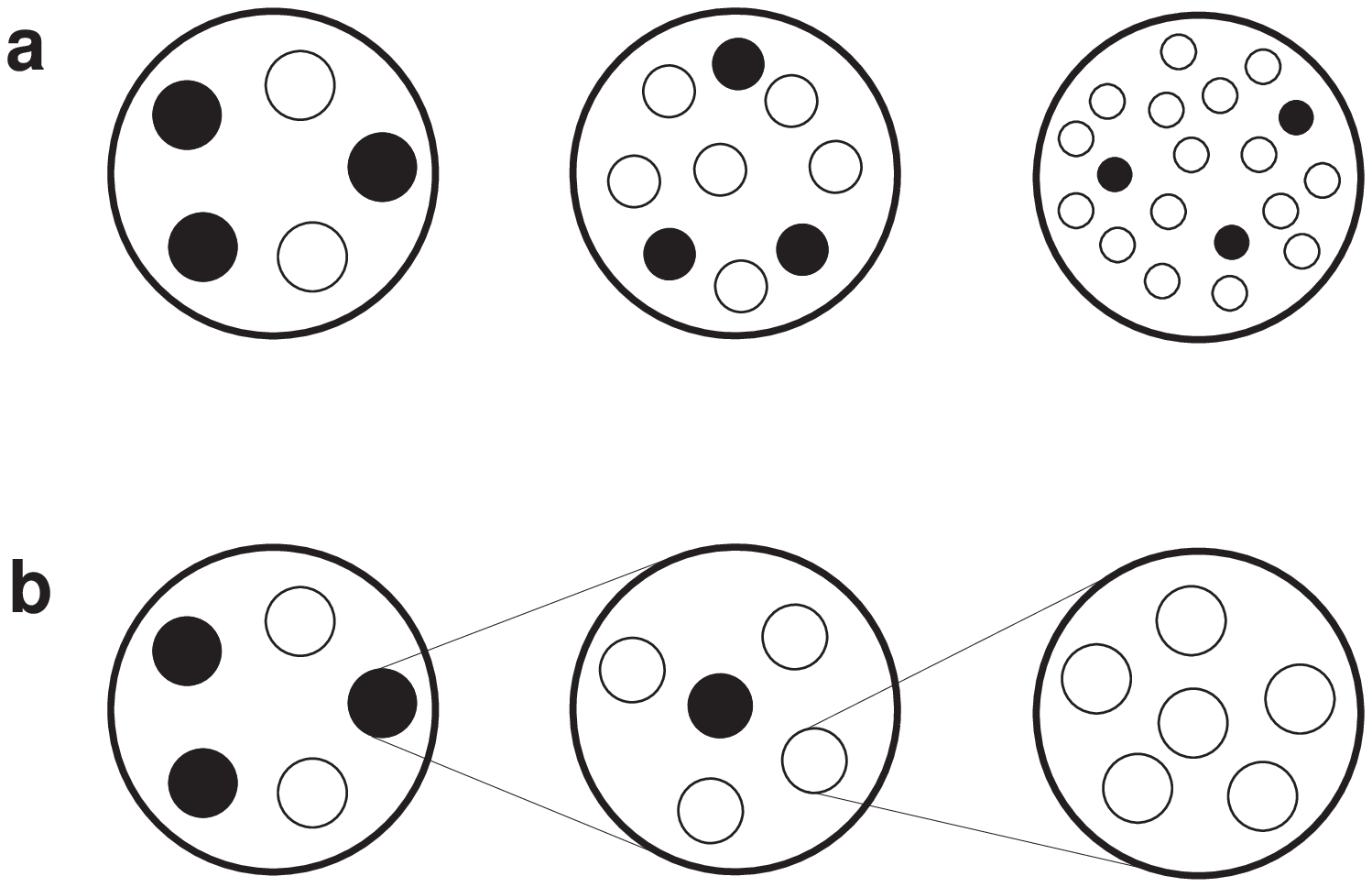,height=7cm}
\end{center}
\caption{Nucleon consisting of the valence and sea quarks - with different resolutions,
 see text.}
\label{fg1}
\end{figure}

2) Nucleon consists of the three quarks and partons (gluons + $q\overline{q}$
pairs) mediating the interaction between them, as sketched in the Fig. \ref
{fg1}{\it a}, where the individual pictures represent some terms in the sum (%
\ref{mss31}). The flavors and spins of all the quarks in each the picture
are mutually cancelled, up to the three quarks giving additively the
corresponding nucleon quantum numbers. These three quarks are in the figure
marked by black and in our approach are identified with the {\it valence
quarks}. The reason, that such identification is quite sensible, is the
following. Apparently, the sum (\ref{mss31}) can be split into quark and
antiquark parts $P_q(\epsilon ),P_{\overline{q}}(\epsilon )$, then our
valence term reads 
\begin{equation}
\label{mssa32}P_{val}(\epsilon )=P_q(\epsilon )-P_{\overline{q}}(\epsilon ), 
\end{equation}
which in the $x-$representation exactly corresponds to the current
definition of the valence quarks. Correspondingly, the unmarked quarks are
identified with the {\it sea quarks}. But both the kinds of quarks have the
same energy distributions $\rho _j(\epsilon )$ entering the Eq. (\ref{mss31}%
), in this sense they are completely equivalent. On the other hand, it is
obvious, that for the valence quarks, in Eq. (\ref{mss31}) only ''black''
quarks from the figure contribute, therefore if $\rho _j$ is assumed in the
first approximation independent on the flavor, then 
\begin{equation}
\label{mss33}P_{val}(\epsilon )=3\sum_{j=3}^\infty r_j\rho _j(\epsilon ). 
\end{equation}

3) The quarks carry only part of the nucleon energy (mass), 
\begin{equation}
\label{mss34}\int P(\epsilon )\epsilon d\epsilon =c_qM, 
\end{equation}
where the factor $c_q$ equals roughly one half, the rest is carried by the
gluons. In the first approximation we shall assume this factor is valid also
for any term in the sum (\ref{mss31}), 
\begin{equation}
\label{mss35}j\int \rho _j(\epsilon )\epsilon d\epsilon =c_qM, 
\end{equation}
which in other words means the ratio of the total energies of quarks and
gluons, together constituting the nucleon mass, is the same for all possible
states sketched in Fig. \ref{fg1}{\it a}.

4) We assume all the quarks in the nucleon state $j$ have approximately the
same effective mass $m_j$ and denote $x_0\equiv m_j/M$. One can expect, for
higher $j$ the parameter $x_0$ will drop and so the sum (\ref{mss33}) can be
substituted by the integral 
\begin{equation}
\label{mss37}P_{val}(\epsilon )=3\int \mu _V(x_0)\rho (\epsilon
,x_0)dx_0,\qquad \int \mu _V(x_0)dx_0=1. 
\end{equation}
Obviously, Eq. (\ref{mss31}) can be with the use of Eq. (\ref{mss35})
rewritten in a similar way: 
\begin{equation}
\label{mss38}P(\epsilon )=\int \mu (x_0)\rho (\epsilon ,x_0)dx_0, 
\end{equation}
where 
\begin{equation}
\label{mssa38}\mu (x_0)=\mu _V(x_0)\frac{c_qM}{\overline{\epsilon }(x_0)}%
,\qquad \overline{\epsilon }(x_0)=\int \rho (\epsilon ,x_0)\epsilon
d\epsilon . 
\end{equation}
The physical meaning of the distributions $\mu _V,\mu $ is the following.
The distribution $\mu (x_0)$ represents a probability, that the effective
mass of the quark, which the probing lepton interacts with, is $x_0$ or
alternatively, $\mu (x_0)dx_0$ is the number of quarks in the interval $%
\left\langle x_0,x_0+dx_0\right\rangle $, which the lepton has chance to
interact with. On the other hand, the (normalized) distribution $\mu _V(x_0)$
can be interpreted as a probability, that the exchanging photon
''distinguishes'' the quarks with the effective mass $x_0$ - as expressed by
the pictures with different granularity in Fig. \ref{fg1}{\it a. } In this
sense, each picture in the Fig. \ref{fg1}{\it a} can be labeled by some $x_0$%
, equally as the corresponding term $\rho (\epsilon ,x_0)$ in the integral (%
\ref{mss37}). Obviously, at the same time the $\mu _V(x_0)$ represents also
the distribution of effective masses corresponding to the valence quark
term. Intuitively, the probability of different contributions in Fig. \ref
{fg1}{\it a} should depend also on{\it \ }$Q^2$ (higher $Q^2=$'better
resolution'), so we expect

\begin{equation}
\label{mss39}\mu (x_0)\rightarrow \mu (x_0,Q^2),\qquad \mu
_V(x_0)\rightarrow \mu _V(x_0,Q^2). 
\end{equation}
In the next, we shall identify these distributions with that suggested on
the end of the previous section.

5) The relations (P3.52), (P3.56) and (P3.20) give the recipe how to obtain
the structure function $F_2$ from a given energy distribution of the partons
with some fixed value $x_0$ and charge $e_q:$%
$$
F_2(x,x_0)=e_q^2\varphi (x,x_0), 
$$
\begin{equation}
\label{mss310}\varphi (x,x_0)=-x^2\int_x^1\frac{2V_0^{\prime }(\xi )\xi }{%
\xi ^2+x_0^2}d\xi ,\qquad 2V_0^{\prime }(\xi )\xi =\pm M\rho (\epsilon
,x_0),\qquad \epsilon =\frac M2\left( \xi +\frac{x_0^2}\xi \right) , 
\end{equation}
where the sign $+(-)$ in the second relation refers to the region $\xi <x_0\
(\xi >x_0).$ For the application of this procedure to Eqs. (\ref{mss37}), (%
\ref{mss38}) one has to weight the contributions integrated over $x_0$ by
the corresponding (mean) charge squared. Apparently, the charge weight of
the valence quarks is the constant 
\begin{equation}
\label{mss311}w_{val}=\frac 13\left[ \left( \frac 23\right) ^2+\left( \frac
23\right) ^2+\left( \frac 13\right) ^2\right] =\frac 13 
\end{equation}
for the proton and similarly for the neutron, $w_{val}=2/9.$ For the sea we
assume in the first approximation the ''equilibrated mixture'' of the quarks 
$u:d:s=1:1:1$, so 
\begin{equation}
\label{mss312}w_{sea}=\frac 13\left[ \left( \frac 23\right) ^2+\left( \frac
13\right) ^2+\left( \frac 13\right) ^2\right] =\frac 29. 
\end{equation}
Then for the nucleon with $j$ quarks we get 
\begin{equation}
\label{mss313}w_j=\frac{3w_{val}+(j-3)w_{sea}}j=w_{sea}+\frac{%
3(w_{val}-w_{sea})}j, 
\end{equation}
or in terms of $x_0$ 
\begin{equation}
\label{mss314}w(x_0)=w_{sea}+3(w_{val}-w_{sea})\frac{\overline{\epsilon }%
(x_0)}{c_qM}. 
\end{equation}
Therefore, the energy distributions (\ref{mss37}), (\ref{mss38}) generate
the corresponding structure functions: 
\begin{equation}
\label{mss315}F_{2val}(x,Q^2)=3w_{val}\int \mu _V(x_0,Q^2)\varphi
(x,x_0)dx_0, 
\end{equation}
\begin{equation}
\label{mss316}F_2(x,Q^2)=\int \left( \frac{c_qM}{\overline{\epsilon }(x_0)}%
w_{sea}+3(w_{val}-w_{sea})\right) \mu _V(x_0,Q^2)\varphi (x,x_0)dx_0. 
\end{equation}

6) Now, let us pay attention to the spin structure functions. According to
the concept suggested in item 2), only valence quarks contribute to the
nucleon spin. First, we shall consider the spin functions generated by the
valence quarks with some fixed effective mass $x_0,$ then we shall easily
proceed to the case with the distribution $\mu _V(x_0)$.

In \cite{zav2}, for sake of simplicity, we have assumed all the three
valence quarks contribute to the proton spin equally [Eq. (Q2.7)]. On the
other hand it is obvious the quark symmetry group can impose an extra
constraint on the contributions of different quark flavors as it follows
e.g. from the philosophy of the well known Bjorken \cite{bjo} and
Ellis-Jaffe \cite{eja} sum rules based on the symmetries {\it U(6)} and {\it %
SU(3)}. In our consideration we shall not strictly assume any particular
group of symmetry, but the different spin contributions of {\it u-} and {\it %
d-}quarks will be expressed by a free parameter $a,\ 0\leq a\leq 1$, having
in the notation of Eq. (Q2.7), e.g. for the proton, the following sense 
\begin{equation}
\label{ms31}\Delta h_u(p_0)=2ah(p_0),\qquad \Delta h_d(p_0)=(1-2a)h(p_0), 
\end{equation}
where $h$ is the valence distribution 
\begin{equation}
\label{ms31a}u(p_0)=d(p_0)\equiv h(p_0),\qquad \int h(p_0)d^3p=1, 
\end{equation}
which is not, due to different normalization, identical with the
distribution $\rho (\epsilon )$, but the both are simply related 
\begin{equation}
\label{ms32}\rho (\epsilon )=4\pi \epsilon \sqrt{\epsilon ^2-m^2}h(\epsilon
), 
\end{equation}
in the same way, as the distributions $P,G$ in Eq. (P3.14).

In the case of proton, there are the particular cases:

\noindent a) $a=0$ corresponds to the mutual spin orientation of the three
valence quarks $(s_u,s_u,s_d)=(-1,+1,+1).$

\noindent b) $a=1/3$ corresponds to the oversimplified scenario studied in 
\cite{zav2}, assuming the equal contribution of all the three quarks; $%
(s_u,s_u,s_d)=(+1/3,+1/3,+1/3).$

\noindent c) $a=2/3$ corresponds to the non-relativistic {\it SU(6)}
approach. From the wave function%
$$
\mid p,\uparrow \rangle =\frac 1{\sqrt{2}}\left( \frac 1{\sqrt{6}}\left|
(ud+du)u-2uud\right\rangle \otimes \frac 1{\sqrt{6}}\left| (\uparrow
\downarrow +\downarrow \uparrow )\uparrow -2\uparrow \uparrow \downarrow
\right\rangle \right. 
$$
\begin{equation}
\label{ms32a}+\frac 1{\sqrt{2}}\left. \left| (ud-du)u\right\rangle \otimes
\frac 1{\sqrt{2}}\left| (\uparrow \downarrow -\downarrow \uparrow )\uparrow
\right\rangle \right) 
\end{equation}
one can easily show the mean value of the spin carried by the $d(u)-$ quarks
is $-1/3(4/3)$, i.e. $(s_u,s_u,s_d)=(+2/3,+2/3,-1/3)$, which agrees with $%
a=2/3$ in Eq. (\ref{ms31}).

\noindent d) $a=1$ corresponds to the mutual orientation of the three quarks 
$(s_u,s_u,s_d)=(+1,+1,-1).$

So, the proton spin function $H(p_0)$ entering the master equation (\ref
{ms21}) and expressed in terms of the functions (\ref{ms31}) reads 
\begin{equation}
\label{ms33}H^p(p_0)=\frac 89au(p_0)+\frac 19(1-2a)d(p_0)=\frac
19(1+6a)h(p_0). 
\end{equation}
Assuming the neutron is isospin symmetric, its corresponding spin function
will be 
\begin{equation}
\label{ms34}H^n(p_0)=\frac 49(1-2a)u(p_0)+\frac 29ad(p_0)=\frac
19(4-6a)h(p_0), 
\end{equation}
therefore the corresponding equations for the nucleon spin structure
functions read 
$$
g_j^p(x,x_0)=w_{spin}^p\psi _j(x,x_0),\qquad g_j^n(x,x_0)=w_{spin}^n\psi
_j(x,x_0),\qquad j=1,2, 
$$
\begin{equation}
\label{ms35}w_{spin}^p=\frac 19(1+6a),\qquad w_{spin}^n=\frac 19(4-6a), 
\end{equation}
where, in accordance with Eqs. (Q2.17), (Q2.18) 
\begin{equation}
\label{ms37}\psi _1(x,x_0)=\frac m2\int \frac{h(p_0)}{p_0}\sqrt{\frac{%
p_0+p_1 }{p_0-p_1}}\delta \left( \frac{p_0+p_1}M-x\right) d^3p, 
\end{equation}
\begin{equation}
\label{ms38}\psi _2(x,x_0)=-\frac m2\int \frac{h(p_0)}{p_0}\frac{p_1}{\sqrt{%
p_0^2-p_1^2}}\delta \left( \frac{p_0+p_1}M-x\right) d^3p. 
\end{equation}
The function $\psi _1(x,x_0)$ can be with the use of Eqs. (Q2.24),(Q2.27)
expanded 
\begin{equation}
\label{ms324}\psi _1(x,x_0)=\frac{x_0}2\sum_{j=0}^\infty {\binom{{-\frac 12} 
}{j}}(-1)^jV_{-j-3/2}(x)\left( \frac x2\right) ^{j+1/2}. 
\end{equation}
Since Eq. (P3.56) implies 
\begin{equation}
\label{ms325}V_{-j-3/2}(x)=-\int_x^1V_0^{\prime }(\xi )\left( \frac{2\xi }{%
\xi ^2+x_0^2}\right) ^{j+3/2}d\xi , 
\end{equation}
one can easily show the sum in Eq. (\ref{ms324}) gives 
\begin{equation}
\label{ms326}\psi _1(x,x_0)=-x_0\int_x^1\frac{V_0^{\prime }(\xi )\xi }{\xi
^2+x_0^2}\sqrt{\frac{x\xi }{\xi ^2+x_0^2-x\xi }}d\xi . 
\end{equation}
Similar manipulation with the function $\psi _2$ gives the result 
\begin{equation}
\label{ms327}\psi _2(x,x_0)=-\frac{x_0}2\int_x^1\frac{V_0^{\prime }(\xi )\xi 
}{\xi ^2+x_0^2}\frac{\xi ^2+x_0^2-2x\xi }{\sqrt{x\xi (\xi ^2+x_0^2-x\xi )}}%
d\xi . 
\end{equation}
Obviously, for the case with the distribution $\mu _V,$ the corresponding
spin structure functions read 
\begin{equation}
\label{ms328}g_j(x,Q^2)=w_{spin}\int \mu _V(x_0,Q^2)\psi
_j(x,x_0)dx_0,\qquad j=1,2. 
\end{equation}
Let us note, the structure functions $F_2,F_{2val},g_1,g_2$ are not
independent, all of them are in the corresponding way generated by the
distributions $\mu _V$ and $V_0$ (or, equivalently by $\rho $).

7) Now, to make the construction suggested above applicable for some
quantitative comparison with the experimental data, we have to use some
reasonable, simple and sufficiently flexible parameterization for the
unknown functions $\mu _V$ and $V_0$. We suggest the following.

\noindent a){\it \ }Normalized distribution $\mu _V$ is assumed in the form 
\begin{equation}
\label{ms329}\mu _V(x_0,Q^2)=\frac{\Gamma (r+s+2)}{\Gamma (r+1)\Gamma (s+1)}%
\cdot x_0^r(1-x_0)^s,\qquad 0<x_0<1, 
\end{equation}
where the $Q^2-$dependence is involved in the parameters $r,s.$

\noindent b){\it \ }For the function $V_0^{\prime }(x)x$ we shall use the
parameterization suggested in Eqs. (Q2.34)-(Q2.38) 
\begin{equation}
\label{ms330}V_0^{\prime }(x)x=\mp c_{norm}\cdot f(x,x_0),\qquad
f(x,x_0)\equiv \left[ (1-x)\left( 1-\frac{x_0^2}x\right) \right] ^{\alpha
/2x_0}, 
\end{equation}
where the upper (lower) sign in the first relation refers to the region $%
x>x_0\quad (x<x_0)$ and 
\begin{equation}
\label{ms331}c_{norm}=\left[ \int_{x_0}^1f(x,x_0)\left( 1-\frac{x_0^2}{x^2}%
\right) dx\right] ^{-1}, 
\end{equation}
which follows e.g. from Eqs. (Q2.34), (P3.25). Now, apparently one has to
accept the parameter $\alpha \approx m/\left\langle E_{kin}\right\rangle $
depends on $x_0$ as well. Let us consider the following. Sequence of the
pictures in Fig. \ref{fg1}{\it a} can be understood as the pictures of the
one and the same nucleon, but ''seen with different resolutions'' as
outlined in Fig. \ref{fg1}{\it b}. Then, it is natural to assume the
four-momentum $P$ of the parton from some picture can be obtained from
four-momenta $p_\lambda $ of $n$ partons in a picture more rightwards,
representing the parton ''seen with better resolution'': 
\begin{equation}
\label{ms332}P=\sum_\lambda p_\lambda . 
\end{equation}
Obviously, the mean values satisfy 
\begin{equation}
\label{ms333}\left\langle P_0\right\rangle =n\left\langle p_{\lambda
0}\right\rangle ,\qquad \left\langle \left| \vec P\right| \right\rangle
=c_{corr}\cdot n\left\langle \left| \vec p_\lambda \right| \right\rangle
,\qquad 0\leq c_{corr}\leq 1, 
\end{equation}
where $c_{corr}$ equals $0(1)$ for the extreme case, when the motion of the
partons in the corresponding subset is completely uncorrelated (correlated).
The last relations imply the effective masses and kinetic energies obey 
\begin{equation}
\label{ms334}m(P)\geq n\cdot m(p),\qquad \left\langle
E_{kin}(P)\right\rangle \leq n\cdot \left\langle E_{kin}(p)\right\rangle , 
\end{equation}
which means the quantity $\alpha $ is a non-decreasing function of $x_0$. In
this moment we know nothing more about this function, in the next section we
shall show a reasonable agreement with the experimental data can be obtained
with 
\begin{equation}
\label{ms335}\alpha (x_0)=\alpha _1\left( -\ln (x_0)\right) ^{-\alpha _2}. 
\end{equation}

Since we parameterize the function $V_0^{\prime }$ rather than the function $%
\rho $, it will be useful the quantity $\overline{\epsilon }(x_0),$ defined
in Eq. (\ref{mss38}) and afterwards entering the important Eq. (\ref{mss316}%
), to express also in terms of $V_0^{\prime }$. Obviously, using Eqs. (\ref
{mss310}) and (\ref{ms330}) one gets 
$$
\overline{\epsilon }(x_0)=\int \rho (\epsilon ,x_0)\epsilon d\epsilon
=-\int_{x_0}^1V_0^{\prime }(\xi )\xi \left( \xi +\frac{x_0^2}\xi \right)
\frac M2\left( 1-\frac{x_0^2}{\xi ^2}\right) d\xi 
$$
\begin{equation}
\label{ms336}=c_{norm}\frac M2\int_{x_0}^1f(\xi ,x_0)\left( \xi -\frac{x_0^4 
}{\xi ^3}\right) d\xi . 
\end{equation}

Now, we can our results shortly summarize. If there are given some values of
the free parameters $c_q,a,r,s,\alpha _1,\alpha _2$, then the corresponding
proton and neutron structure functions can be directly calculated according
to Eqs. (\ref{mss315}), (\ref{mss316}), (\ref{ms328}), where the
distribution $\mu _V$ is given by Eq. (\ref{ms329}), the function $\overline{%
\epsilon }(x_0)$ by Eq. (\ref{ms336}) with the use of Eqs. (\ref{ms330}), (%
\ref{ms331}), (\ref{ms335}) and the functions $\varphi ,\psi _1,\psi _2$
are: 
\begin{equation}
\label{ms337}\varphi (x,x_0)=2x^2\int_x^1\eta (\xi ,x_0)d\xi , 
\end{equation}
\begin{equation}
\label{ms338}\psi _1(x,x_0)=x_0\int_x^1\eta (\xi ,x_0)\sqrt{\frac{x\xi }{\xi
^2+x_0^2-x\xi }}d\xi , 
\end{equation}
\begin{equation}
\label{ms339}\psi _2(x,x_0)=\frac{x_0}2\int_x^1\eta (\xi ,x_0)\frac{\xi
^2+x_0^2-2x\xi }{\sqrt{x\xi (\xi ^2+x_0^2-x\xi )}}d\xi , 
\end{equation}
where 
\begin{equation}
\label{ms340}\eta (\xi ,x_0)=c_{norm}\theta (\xi -x_0)\frac{f(\xi ,x_0)}{\xi
^2+x_0^2}. 
\end{equation}
The last expression is calculated from Eqs. (\ref{ms330}), (\ref{ms331}) and
(\ref{ms335}) with the use of the step function $\theta (y)=+1(-1)$ for $%
y>0\ (y<0).$

\section{Comparison with the experimental data}

\label{sec4}

\setcounter{equation}{0}

Now we shall try to compare our formulas for the structure functions with
the existing data. We shall not attempt to make a consistent, global fit of
the free parameters based on some rigorous fitting procedure, but only show
the set of optimal parameters obtained by their tentative varying on the
computer ''by hand''. Moreover, our constraint will be only agreement with
the proton structure functions $F_2$ and $g_1$. It means that the parameter $%
a,$ controlling asymmetry between the proton and neutron spin functions,
must be somehow fixed before the fitting. For the first approximation we use
the $SU(6)$ value, $a=2/3$ [see item 6c) in the previous section].

\begin{figure}
\begin{center}
\epsfig{file=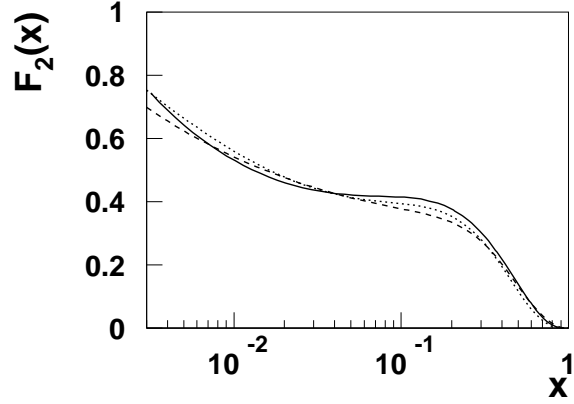,height=7cm}
\end{center}
\caption{Proton structure function $F_2 (x)$ at $Q^2 =10 GeV^2 $. 
The dotted and dashed curves represent the fits of the experimental 
data suggested in {\protect\cite{f2h1}} and {\protect\cite{smc}}. 
The full curve is the 
result of our calculation.}
\label{fg3}
\end{figure}

For a comparison with $F_2$ we use the parameterizations of the world data
suggested in \cite{f2h1} and \cite{smc}, both taken for $Q^2=10GeV^2$. The
data for $g_1$ are taken over from the recent paper \cite{smc} of the SMC
Collaboration. After some checking on the computer, the optimal set of the
free parameters is considered: 
\begin{equation}
\label{ms41}c_q=0.43,\quad r=-0.49,\quad s=6.5,\quad \alpha _1=1.6,\quad
\alpha _2=1.5 
\end{equation}
Results of the calculation of the proton structure functions $g_1$ and $F_2$
with these parameters are shown in Figs. \ref{fg2}, \ref{fg3} together with
the data. Let us remark, the experimental points for $g_1$ correspond to the
values evolved in \cite{smc} to $Q^2=10GeV^2$. In the error bars all the
quoted errors (statistical, systematic and those due to uncertainty of QCD
evolution) are combined. Obviously, the agreement with the experimental data
in both the figures can be considered very good, particularly if we take
into account that our parameterization of the unknown distributions is
perhaps the simplest possible and moreover, the parameters (\ref{ms41})
still may not be optimal.

\begin{figure}
\begin{center}
\epsfig{file=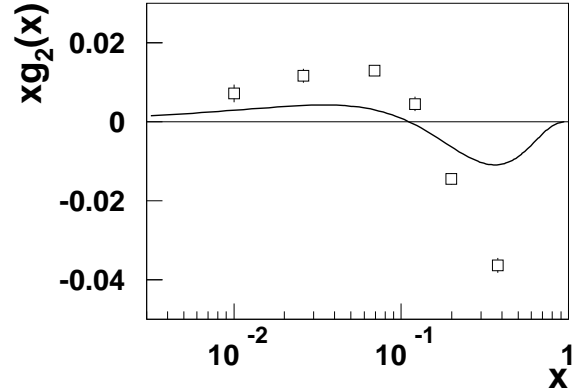,height=7cm}
\end{center}
\caption{Proton structure function $xg_2 (x)$ at $Q^2 =10 GeV^2 $. 
The points represent the term $xg_2 ^{WW} (x)$ (see text), 
the curve is the 
result of our calculation.}
\label{fg4}
\end{figure}

\begin{figure}
\begin{center}
\epsfig{file=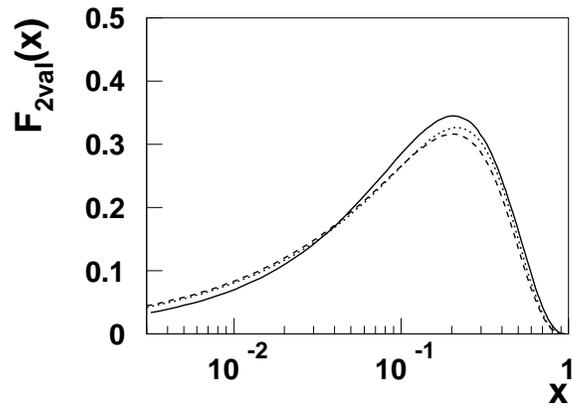,height=7cm}
\end{center}
\caption{Proton structure function $F_{2val} (x)$. 
The dashed and dotted curves represent the functions based on the 
standard global analysis according to the relations (\ref{ms42}) 
and (\ref{ms44}), see text. 
The full curve is the 
result of our calculation.}
\label{fg5}
\end{figure}

Now, having ''tuned'' the free parameters by the $g_1$ and $F_2$, one can
predict the remaining functions $g_2$ and $F_{2val}$. The results are shown
in Figs. \ref{fg4}, \ref{fg5}. Our $xg_2$ surely does not contradict the
experimental data \cite{smcd}, which are compatible with zero - with
statistical errors substantially bigger, than the vertical range of the
figure. But instead of the data, the comparison is done with Wandzura
Wilczek \cite{ww} twist-2 term for $xg_2^{WW}$, which is evaluated in \cite
{smcd} from the corresponding $g_1$. It is obvious, that two completely
different approaches give at least qualitatively very similar results. The
proton valence function $F_{2val}$ in Fig. \ref{fg5} is compared with the
corresponding combination of the distributions $xu_V(x)$ and $xd_V(x)$
obtained (for $Q^2=4GeV$) in \cite{msr} by the standard global analysis: 
\begin{equation}
\label{ms42}F_{2val}(x)=\frac 89xu_V(x)+\frac 19xd_V(x),\quad \int
u_V(x)dx=\int d_V(x)dx=1.\ 
\end{equation}
Apparently, the agreement can be considered good. Nevertheless, let us note,
a part of the difference between both the curves can be due to our
simplification: $u_V\approx d_V$. On the other hand, the distributions of
the $u$- and $d$-valence quarks are not equal and enter the functions $g_1$
and $F_{2val}$ with different weights, see Eqs. (\ref{ms33}), (\ref{ms42}).
At the same time, our prediction for $F_{2val}$ is based on the combination
valid for $g_1$%
\begin{equation}
\label{ms43}a\frac 89u_V(x)+(1-2a)\frac 19d_V(x)\approx \frac{1+6a}9q_V(x),
\end{equation}
which is related to $F_{2val}$%
$$
F_{2val}(x)\approx \frac 89xq_V(x)+\frac 19xq_V(x)=xq_V\approx \frac
9{1+6a}\left( \frac 89xu_V(x)+\frac 19xd_V(x)\right)  
$$
\begin{equation}
\label{ms44}=\frac{16}{15}xu_V(x)-\frac 1{15}xd_V(x)
\end{equation}
for our value $a=2/3$. The last combination is also shown in the Fig. \ref
{fg5} and apparently gives a slightly better agreement with our simplified $%
F_{2val}.$ In any case, one can note, that the two different procedures, the
standard one (uses input on $F_2,F_3^{\nu N}$ + QCD) and our (uses input on $%
F_2,g_1$ + our model) give a very similar picture of the function $%
F_{2val}(x)$, which is not directly measurable.

\section{Discussion}

\label{sec5}

\setcounter{equation}{0}

\begin{figure}
\begin{center}
\epsfig{file=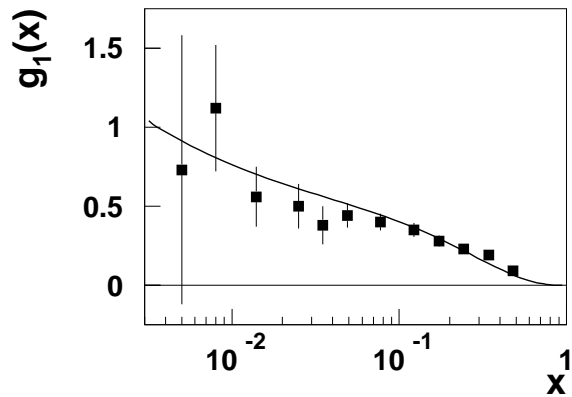,height=7cm}
\end{center}
\caption{Proton spin structure function $g_1 (x)$ at $Q^2 =10 GeV^2 $.
 The points represent experimental data {\protect\cite{smc}}, the curve is the 
result of our calculation.}
\label{fg2}
\end{figure}

Let us make a few comments on the obtained results. First of all, it should
be pointed out, that our structure functions in Figs. \ref{fg2}-\ref{fg5}
are calculated on the basis of very simple parameterization of the unknown
distributions $\mu (x_0)$ and $V_0(x,x_0)$, but on the other hand it is
essential, that the contributions from the individual components of the
quark distribution correctly take into account the intrinsic quark motion,
which is particularly important for the spin structure function. The effect
of this motion on $g_1$ is demonstrated in Fig. Q1 and the fact, that we
succeeded to achieve a good agreement with the data also in Fig. \ref{fg2}
is just thanks to this effect. For a better insight, how our structure
functions are generated, in the Fig. \ref{fg6} we have displayed the initial
distribution function $V_0(x,x_0)$ drawn for a few values $x_0,$ together
with the corresponding structure functions $F_2,g_1,xg_2$. The complete
structure functions are their superpositions - weighted in the corresponding
way with the use of the distribution $\mu _V(x_0)$.

Further, also some other assumptions of the model are possibly
oversimplified, for a more precise calculation, at least some of them could
be rightly modified - but at a price of introducing the additional free
parameters. For example, the constant $w_{sea}$ [see Eq. (\ref{mss312})]
should take into account some suppression of the $s-$ quarks \cite{msr} and
probably should allow a weak dependence on $x_0$. Also for the constant $c_q$
[see Eq. (\ref{mss34})] some $x_0-$dependence should be allowed. Concerning
this constant, let us make one more comment. The standard global fit \cite
{msr} suggests (at $Q^2=10GeV^2$) the quarks carry $\simeq 56\%$ of the
nucleon energy and our fitted value $c_q$ from the Eq. (\ref{mss34}) is
roughly $43\%$. This difference is mainly due to the different relations
between the distribution and structure function in both the approaches, see
Eqs. (P3.38), (P3.59). The second relation (valid for a subset of quarks
with effective mass $x_0$), multiplied by $x^2$ and then integrated by parts
gives 
\begin{equation}
\label{ms51}\int_{x_0^2}^1xF(x,x_0)dx=\frac 14\int_{x_0^2}^1F_2(x,x_0)\left(
3+\frac{x_0^2}{x^2}\right) dx, 
\end{equation}
which for the static quarks [$F(x,x_0)\simeq F_2(x,x_0)/x\simeq \delta
(x-x_0)$, see discussion after Eq. (P3.59)] coincides with the standard
relation. Nevertheless, generally both the relations imply different rate of
the nucleon energy carried by quarks. One can check numerically that for our 
$F_2(x,x_0)$ in a dominant region of $x_0$ the term $(x/x_0)^2$ in the
integral (\ref{ms51}) plays a minor role (see also Fig. \ref{fg6}, positions
of the maxima of $F_2$'s are above the corresponding $x_0$, in particular
for lower $x_0$), so as a result we get $3/4$ of the standard estimation of
the quark contribution to the nucleon energy. This ratio agrees with the
ratio obtained from the corresponding fits: $3/4\simeq 43\%\ /\ 56\%.$

In the previous section we mentioned the effect of different shape of the $u-
$ and $d-$quark distributions. Also the result $g_1^n(x)=0$ following from
Eq. (\ref{ms34}) is a consequence of the simplifying assumptions $%
u^n=d^n=u^p=d^p\equiv h$ and $a=2/3$. A more general approach $u\neq d$
would give
\begin{equation}
\label{msaa51}H^n(p_0)=\frac 4{27}\left( -u^n(p_0)+d^n(p_0)\right) =\frac
4{27}\left( -d^p(p_0)+u^p(p_0)\right) ,
\end{equation}
so $u\neq d$ implies $H^n\neq 0$. In fact, the global fit analysis proves,
that $d_{val}^p(x)$ is slightly ''narrower'' than $u_{val}^p(x).$ It means,
considering very qualitatively, in accordance with the last equation, in the
function $g_1^n(x)$ the negative term should dominate for smaller $x$, which
does not contradict the experimental data. Obviously, a proper accounting
for the difference $u\neq d$ into the model should enable to calculate
consistently in a better approximation not only the proton and neutron
structure functions $F_2,g_1,g_2$, but also the neutrino structure
functions. Apparently then one could make a ''super-global'' fit covering
the both unpolarized and polarized DIS data. As a result, the
flavor-dependent quark distributions $V_0(x,x_0)$ [or equivalently $\rho
(\epsilon ,x_0)$] together with the corresponding effective mass
distributions and the parameter $a$ controlling the relative spin
contribution of the $u-$ and $d-$quarks, could be obtained.

Finally, let us point out, inclusion the spin structure function into the
fit in our model enables to obtain some information about the distribution
of the quark effective masses. Within our approach there are two
distributions, $\mu _V$ and $\mu $, relevant for the description of the
quark effective masses in the nucleon. The extrapolation of our
parameterization for the $\mu $ distribution with the use of the relations $%
\alpha \approx m/\left\langle E_{kin}\right\rangle $ and (\ref{mssa38}),(\ref
{ms329}),(\ref{ms335}),(\ref{ms41}) give for $x_0\rightarrow 0:$%
\begin{equation}
\label{msa51}\mu (x_0)\sim \frac{\mu (x_0)}{\overline{\epsilon }(x_0)}%
\rightarrow \frac{x_0^r}{Mx_0+\left\langle E_{kin}\right\rangle }\rightarrow 
\frac{x_0^{-1.49}}{\left| \ln x_0\right| ^{1.5}}, 
\end{equation}
which implies the extrapolated $\mu $ is not integrable in this limit. On
the other hand, the basic distribution $\mu _V$, parameterized by Eq. (\ref
{ms329}) with the $r,s$ from the set (\ref{ms41}) and with the use of the
known relation $z\Gamma (z)=\Gamma (z+1)$ can give an estimate of the mean
value: 
\begin{equation}
\label{ms52}\left\langle x_0\right\rangle _V=\frac{r+1}{r+s+2}\simeq 0.064, 
\end{equation}
i.e. $\left\langle m\right\rangle \simeq 60MeV$ for $Q^2=10GeV^2$. The
corresponding kinetic term calculated as 
\begin{equation}
\label{msa52}\left\langle E_{kin}\right\rangle _V=\int \mu _V(x_0)(\overline{%
\epsilon }(x_0)-Mx_0)dx_0 
\end{equation}
gives a similar number ($\simeq 60MeV$). Let us recall, these numbers are
related to the valence quark term. The $Q^2$-dependence is involved only in
the distribution $\mu _V(x_0,Q^2)$, i.e. in our parameterization (\ref{ms329}%
) only via the powers $r(Q^2),s(Q^2)$. It follows, the structure functions,
which enhance in a low$-x$ region for increasing $Q^2$, must be generated by
the distribution $\mu _V(x_0,Q^2)$ in which the mean effective mass $%
\left\langle x_0\right\rangle _V$ drops for increasing $Q^2$ - in the
qualitative agreement with an intuitive expectation.

\section{Summary}

\label{sec6}

In the present paper, with the use of the results obtained in the preceding
ones \cite{zav1},\cite{zav2}, we proposed an alternative covariant
formulation of the QPM. The initial postulates of the standard and our
approach are basically the same, despite of that the relations between the
structure and distribution functions obtained in both the approaches are not
identical. It is due to the fact, that in the standard approach the
intrinsic quark motion is effectively suppressed by the use of the
approximation $p_\mu =xP_\mu $. On the other hand, we have shown the master
equations can be solved without the use of the this approximation, so in the
corresponding solution the quark intrinsic motion is consistently taken into
account. On the basis of the obtained relations (a priori valid for the
version of naive QPM - with non-static quarks on mass shell) we propose the
model, in which the distributions $(\mu _V,V_0)$ reflecting the parton
dynamics are introduced with some free parameters. With the use of this
model we calculated simultaneously the proton structure functions $%
F_2,F_{2val},g_1,g_2$, assuming only the valence quarks term contributes to
the proton spin. By a comparison with the data on $F_2,g_1$ $(Q^2=10GeV^2)$
we fix the free parameters and then on the basis of these parameters the
functions $F_{2val}$ and $g_2$ are predicted. We found out:

1) Both the unpolarized structure functions are well reproduced by the
model. The comparison is done with the data on $F_2$ and with the $F_{2val}$
obtained from the standard global analysis data.

2) At the same time, the model well agrees with the data on $g_1$. The
calculated $g_2$ does not contradict the existing experimental data as well,
but since in the data there are still rather big statistical errors, it is
hard to say more.

3) Analysis of the fixed parameters within our approach suggests:

\noindent $i)$ The quarks carry less the proton energy (almost by the factor
3/4), than estimated from the standard analysis.

\noindent $ii)$ The average effective mass related to the valence quark term
can be roughly $60MeV$ and a similar energy can be ascribed to the
corresponding motion.

So on the end we can underline the proposed model offers apart of the other
results also the consistent explanation, why the experimentally measured
proton spin function $g_1$ is less, than it is predicted in the standard
approach based on the QCD improved QPM.

\section{Appendix}

\setcounter{equation}{0}

\begin{figure}
\begin{center}
\epsfig{file=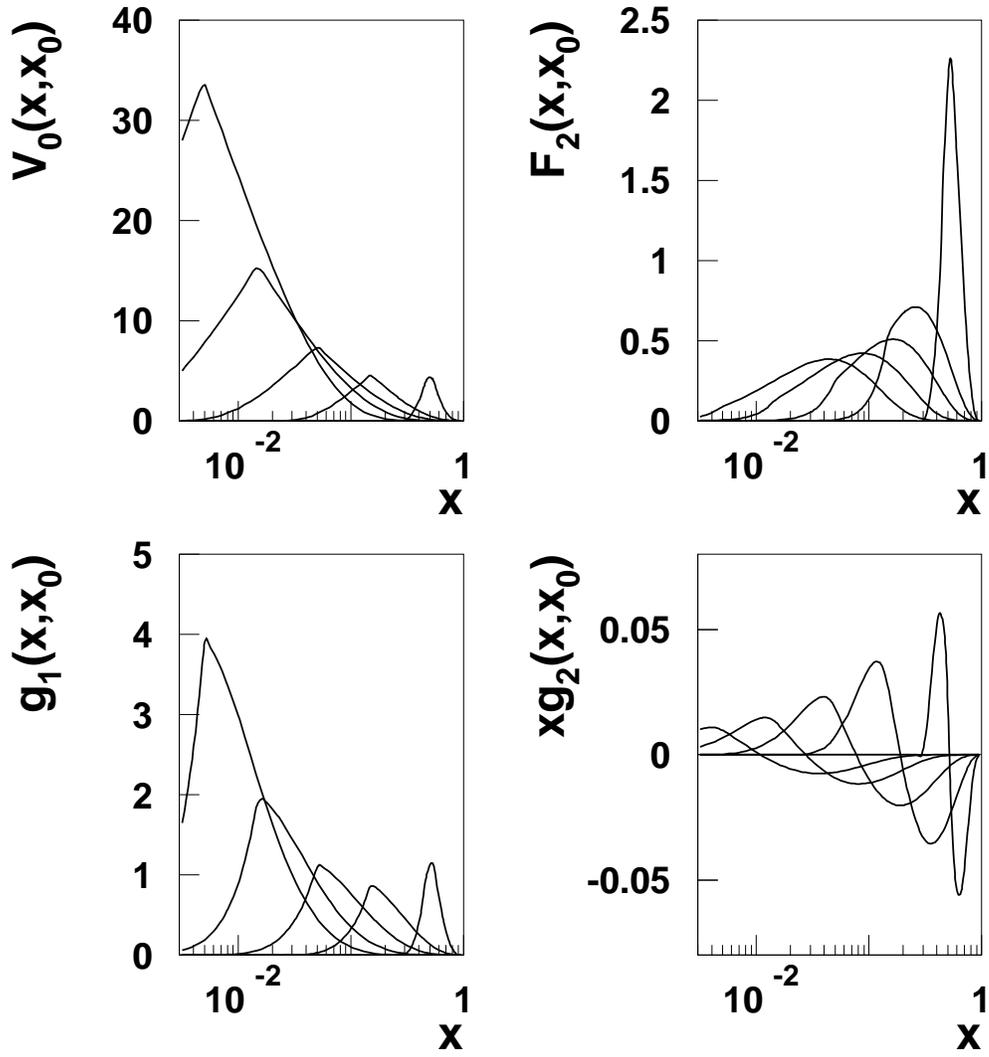,height=16cm}
\end{center}
\caption{Distribution functions $V_0 (x,x_0)$ drawn for 
$x_0 =0.005,0.015,0.05,0.15,0.5$ and the structure functions 
generated correspondingly. The calculation is based on the Eqs. 
(\ref{ms330}), (\ref{mss310}), (\ref{ms326}) and (\ref{ms327}).}
\label{fg6}
\end{figure}

\begin{figure}
\begin{center}
\epsfig{file=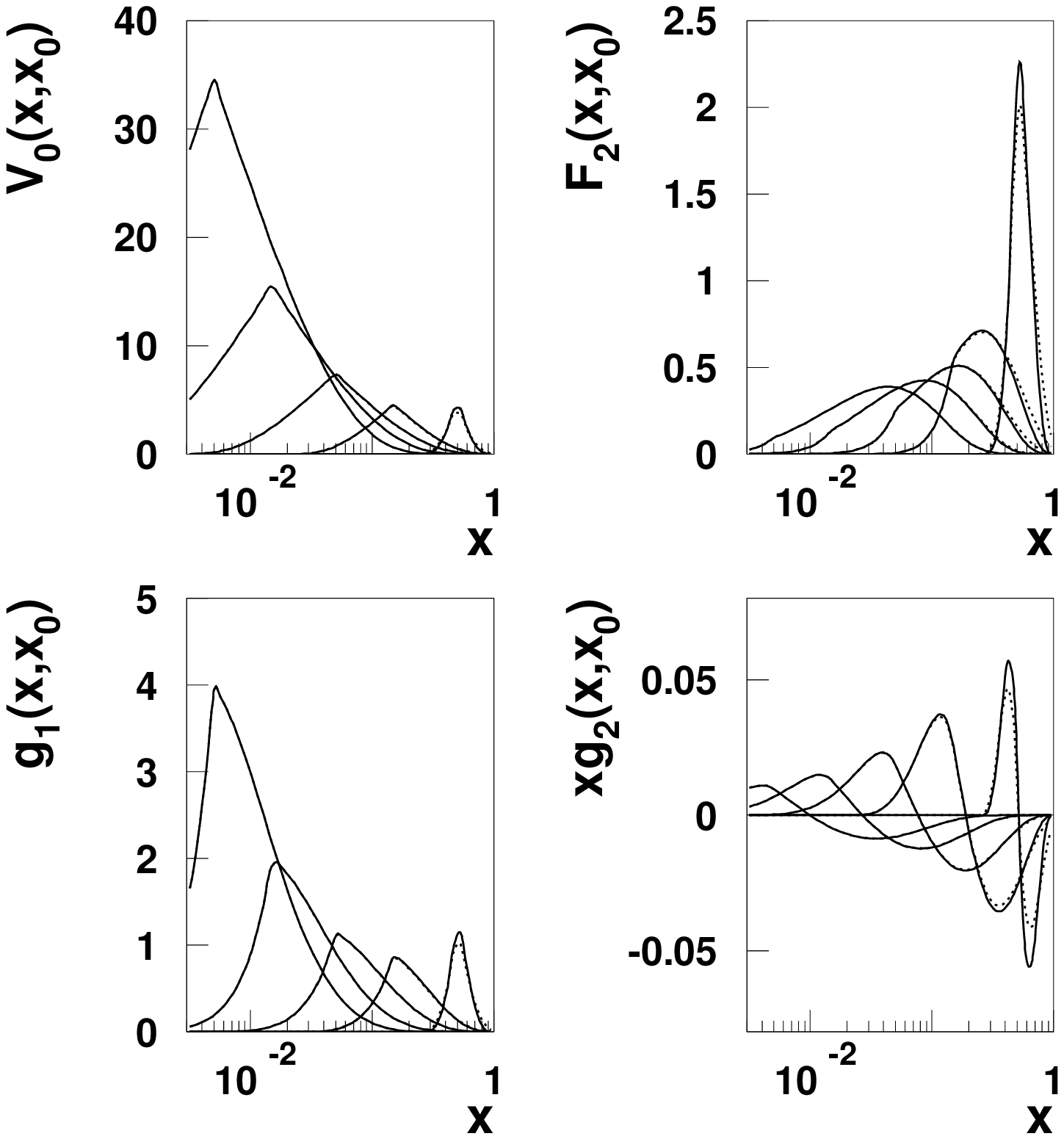,height=16cm}
\end{center}
\caption{Distribution functions $V_0 (x,x_0)$ and the corresponding
structure functions drawn for 
$x_0 =0.005,0.015,0.05,0.15,0.5$. The functions are calculated according
to Eqs. (\ref{app29}) - (\ref{app32}) for $Q^2\rightarrow \infty $ (full 
lines) and $Q^2 = 3GeV^2/c^2$ (dotted lines).}
\label{fg7}
\end{figure}

In this appendix we shall explain in more detail, that approximation (\ref
{app0}) is sufficiently correct for the purpose of this paper. Structure
functions given by Eqs. (P3.52), (P3.47) and (Q2.13), (Q2.14) are given by
the integrals of the form 
\begin{equation}
\label{app1}\int K(p_0,p_1)\delta \left( \frac{pq}{M\nu }-x\right)
d^3p,\qquad p_0^2=p_1^2+p_2^2+p_3^2+m^2,\qquad x=\frac{Q^2}{2M\nu }. 
\end{equation}
In accordance with Eqs. (P2.8)-(P2.13) one can express 
\begin{equation}
\label{app2}\frac{pq}{M\nu }=\frac{p_0+p_1}M+\frac{p_1}{k_0}x-\frac{p_T}M 
\sqrt{\left( \frac{4M^2}{Q^2}-\frac{M^2}{k_0^2}\right) x^2-\frac{2M}{k_0}x}%
\cos \varphi 
\end{equation}
and if one assumes $k_0\gg M,$ then the second term can be neglected and
after introducing the variables $p_T,\varphi $ 
\begin{equation}
\label{app3}p_2=p_T\cos \varphi ,\qquad p_3=p_T\sin \varphi ;\qquad
d^3p=dp_1p_Tdp_Td\varphi 
\end{equation}
the integral (\ref{app1}) reads 
\begin{equation}
\label{app4}I(x,a)=\int K(p_0,p_1)\delta \left( \frac{p_0+p_1}M-ap_T\cos
\varphi -x\right) dp_1p_Tdp_Td\varphi , 
\end{equation}
where 
\begin{equation}
\label{app5}a=\frac 1M\sqrt{\left( \frac{4M^2}{Q^2}-\frac{M^2}{k_0^2}\right)
x^2-\frac{2M}{k_0}x}. 
\end{equation}
Obviously, our approximation (\ref{app0}) used in Eqs. (P3.53) and (Q2.15)
corresponds to $a=0$ i.e. $Q^2\rightarrow \infty $ and now we shall try to
estimate the effect of the finite $Q^2$ in Eq. (\ref{app4}). First, we shall
integrate over $p_1$ using the rule 
\begin{equation}
\label{app6}\delta \left( f(x)\right) dx=\sum_j\frac{\delta \left(
x-x_j\right) }{f^{\prime }\left( x_j\right) }dx,\qquad f\left( x_j\right)
=0, 
\end{equation}
which with (just one $p_1$) root in the argument of the $\delta -$function
in Eq. (\ref{app4}) gives 
\begin{equation}
\label{app7}\delta \left( \frac{p_0+p_1}M-ap_T\cos \varphi -x\right) dp_1= 
\frac{{\bf p}{\rm _0}}{x+ap_T\cos \varphi }\delta \left( p_1-{\bf p}{\rm _1}%
\right) dp_1, 
\end{equation}
where 
\begin{equation}
\label{appe7}{\bf p}{\rm _0}=\frac 12\left[ M\left( x+ap_T\cos \varphi
\right) +\frac{m^2+p_T^2}{M\left( x+ap_T\cos \varphi \right) }\right] , 
\end{equation}
\begin{equation}
\label{app8}{\bf p}{\rm _1}=\frac 12\left[ M\left( x+ap_T\cos \varphi
\right) -\frac{m^2+p_T^2}{M\left( x+ap_T\cos \varphi \right) }\right] . 
\end{equation}
Now the integral (\ref{app4}) can be expressed 
\begin{equation}
\label{app9}I(x,a)=\int K({\bf p}{\rm _0,{\bf p}_1})\frac{{\bf p}{\rm _0}}{%
x+ap_T\cos \varphi }p_Tdp_Td\varphi . 
\end{equation}
The upper limit of $p_T$ can be with the use of Eq. (P3.11) calculated from
the equation 
\begin{equation}
\label{app10}{\bf p}{\rm _0}=E_{\max }=\frac M2\left( 1+x_0^2\right) ,\qquad
x_0=m/M, 
\end{equation}
which implies the equation 
\begin{equation}
\label{app11}Ap_T^2+Bp_T+C=0, 
\end{equation}
where 
\begin{equation}
\label{app12}A=1+M^2a^2\cos {}^2\varphi ,\quad B=2Ma\cos \varphi (xM-E_{\max
}),\quad C=-M^2(1-x)(x-x_0^2) 
\end{equation}
and correspondingly 
\begin{equation}
\label{app13}p_{T\max }=\frac{-B\pm \sqrt{B^2-4AC}}{2A}. 
\end{equation}
Since one requires $p_T\geq 0$ and obviously we have $A>0,$ $C\leq 0$ in our
region $x_0^2\leq x\leq 1,$ so only the upper sign root is acceptable. For $%
Q^2\rightarrow \infty $ this root reads 
\begin{equation}
\label{app14}p_{T\max }=M\sqrt{(1-x)(x-x_0^2)} 
\end{equation}
in accordance with Eq. (P2.19). The integral (\ref{app9}) can be now
expressed 
\begin{equation}
\label{app15}I(x,a)=\int_0^{2\pi }\int_0^{p_{T\max }(x,a\cos \varphi
)}L\left( x+ap_T\cos \varphi ,p_T\right) dp_Td\varphi , 
\end{equation}
where 
\begin{equation}
\label{app16}L\left( x+ap_T\cos \varphi ,p_T\right) =K({\bf p}{\rm _0,{\bf p}%
_1})\frac{{\bf p}{\rm _0}p_T}{x+ap_T\cos \varphi }. 
\end{equation}
Now, using the rule for differentiation of the integral 
\begin{equation}
\label{app17}f(t)=\int_0^{g(t)}h(z,t)dz\Rightarrow \frac{df}{dt}%
=\int_0^{g(t)}\frac{\partial h(z,t)}{\partial t}dz+h\left( g(t),t\right) 
\frac{dg}{dt} 
\end{equation}
and taking into account our integrated function vanishes at the boundary of
phase space, $L(x+ap_{T\max }\cos \varphi ,p_{T\max })=0,$ then the second
term in the last equation equals zero and the integral (\ref{app15}) can be
expanded 
\begin{equation}
\label{app18}I(x,a)=\sum_{j=0}^\infty \left. \frac{\partial ^jI(x,a)}{%
\partial a^j}\right| _{a=0}\frac{a^j}{j!}, 
\end{equation}
where 
\begin{equation}
\label{app19}\left. \frac{\partial ^jI(x,a)}{\partial a^j}\right|
_{a=0}=\int_0^{2\pi }\int_0^{p_{T\max }(x,a\cos \varphi )}\left. \frac{%
\partial ^jL\left( x+ap_T\cos \varphi ,p_T\right) }{\partial a^j}\right|
_{a=0}dp_Td\varphi . 
\end{equation}
At the same time, one can write 
\begin{equation}
\label{app20}\int_0^{2\pi }\int_0^{p_{T\max }(x,a\cos \varphi )}\left. \frac{%
\partial ^jL}{\partial a^j}\right| _{a=0}dp_Td\varphi =\int_0^{2\pi
}\int_0^{p_{T\max }(x,a\cos \varphi )}\left. \frac{\partial ^jL}{\partial x^j%
}\right| _{a=0}\left( p_T\cos \varphi \right) ^jdp_Td\varphi 
\end{equation}
i.e. the $j-$odd terms vanish, so the difference $I(x,a)-I(x,0)$ can be
estimated by an expression of the magnitude of the second order in the
parameter $a.$ Quantitatively, this magnitude depends on the actual shape of
the function $K.$ So let us now calculate the integral (\ref{app15}) for the
functions $V_0,F_2,g_1,g_2$ in accordance with the parameterization used in
this and the previous paper. First of all, one has to replace the
parameterized distribution function $V_0(x)dx$ by the corresponding $%
G(p_0)d^3p.$ The equations (Q2.34), (Q2.37) and (Q2.38) imply 
\begin{equation}
\label{app21}P(p_0)\sim \left[ (1-x)\left( 1-\frac{x_0^2}x\right) \right]
^{\alpha /2x_0},\quad p_0=\frac M2\left( 1+\frac{x_0^2}x\right) , 
\end{equation}
which gives 
\begin{equation}
\label{app22}P(p_0)\sim \left( 1+x_0^2-\frac{2p_0}M\right) ^{\alpha
/2x_0}\sim \left( 1-\frac{p_0}{E_{\max }}\right) ^{\alpha /2x_0},\qquad
E_{\max }=\frac M2\left( 1+x_0^2\right) .\qquad 
\end{equation}
Normalization factor is calculated by integration 
\begin{equation}
\label{app23}N=\int_m^{E_{\max }}\left( 1-\frac{p_0}{E_{\max }}\right)
^{\alpha /2x_0}dp_0=\frac M2\frac{1+x_0^2}{\alpha /2x_0+1}\left( \frac{%
(1-x_0)^2}{1+x_0^2}\right) ^{\alpha /2x_0+1}. 
\end{equation}
The distribution $P(p_0)dp_0$ is related to the distribution $G(p_0)d^3p$ by
Eq. (P3.14), so we obtain 
\begin{equation}
\label{app24}G(p_0)=\frac 1N\frac{(1-p_0/E_{\max })^{\alpha /2x_0}}{4\pi
p_0m \sqrt{(p_0/m)^2-1}}, 
\end{equation}
where $\alpha $ is for the numerical calculation parameterized in accordance
with Eqs. (\ref{ms335}) and (\ref{ms41}). Now, the integrals corresponding
to the functions $V_0,F_2,g_1,g_2$ could be calculated without approximation
(\ref{app0}). In an accordance with Eqs. (P3.47), (P3.52), (Q2.17), (Q2.18)
one has to compute 
\begin{equation}
\label{app25}V_0(x,x_0)=\int G(p_0)\delta \left( \frac{pq}{M\nu }-x\right)
d^3p, 
\end{equation}
\begin{equation}
\label{app26}F_2(x,x_0)=x^2M\int G(p_0)\delta \left( \frac{pq}{M\nu }%
-x\right) \frac{d^3p}{p_0}, 
\end{equation}
\begin{equation}
\label{app27}g_1(x,x_0)=w_{spin}\frac m2\int G(p_0)\frac{p_0+p_1}{\sqrt{%
p_0^2-p_1^2}}\delta \left( \frac{pq}{M\nu }-x\right) \frac{d^3p}{p_0}, 
\end{equation}
\begin{equation}
\label{app28}g_2(x,x_0)=-w_{spin}\frac m2\int G(p_0)\frac{p_1}{\sqrt{%
p_0^2-p_1^2}}\delta \left( \frac{pq}{M\nu }-x\right) \frac{d^3p}{p_0}, 
\end{equation}
where $w_{spin}$ is the factor given by Eq. (\ref{ms35}). The last integrals
are the special cases of the integral (\ref{app1}), so they can be rewritten
in the form of Eq. (\ref{app15}) allowing an easy numerical calculation 
\begin{equation}
\label{app29}V_0(x,x_0)=\int_0^{2\pi }\int_0^{p_{T\max }(x,a\cos \varphi )}G(%
{\bf p}{\rm _0})\frac{{\bf p}{\rm _0}p_T}{x+ap_T\cos \varphi }dp_Td\varphi , 
\end{equation}
\begin{equation}
\label{app30}F_2(x,x_0)=x^2M\int_0^{2\pi }\int_0^{p_{T\max }(x,a\cos \varphi
)}G({\bf p}{\rm _0})\frac{p_T}{x+ap_T\cos \varphi }dp_Td\varphi , 
\end{equation}
\begin{equation}
\label{app31}g_1(x,x_0)=w_{spin}\frac m2\int_0^{2\pi }\int_0^{p_{T\max
}(x,a\cos \varphi )}G({\bf p}{\rm _0})\frac{{\bf p}{\rm _0}+{\bf p}{\rm _1}}{
\sqrt{{\bf p}{\rm _0^2}-{\bf p}{\rm _1^2}}}\frac{p_T}{x+ap_T\cos \varphi }%
dp_Td\varphi , 
\end{equation}
\begin{equation}
\label{app32}g_2(x,x_0)=-w_{spin}\frac m2\int_0^{2\pi }\int_0^{p_{T\max
}(x,a\cos \varphi )}G({\bf p}{\rm _0})\frac{{\bf p}{\rm _1}}{\sqrt{{\bf p}%
{\rm _0^2}-{\bf p}{\rm _1^2}}}\frac{p_T}{x+ap_T\cos \varphi }dp_Td\varphi , 
\end{equation}
where quantities $G,$ $a,$ ${\bf p}_0,$ ${\bf p}_1,$ $p_{T\max }$
are given by Eqs. (\ref{app24}), (\ref{app5}), (\ref{appe7}), (\ref{app8}), (%
\ref{app13}) respectively. These functions $F_2,g_1,g_2$ could replace the
functions (\ref{mss310}), (\ref{ms35}) used in the present paper. In the
Fig. \ref{fg7} we have shown the results of numerical calculation of the
functions (\ref{app29}) - (\ref{app32}) for $a=0,$ which is equivalent to
the approximation (\ref{app0}) and for $a>0$, corresponding to $%
Q^2=3GeV^2/c^2$ and $k_0\rightarrow \infty $. The curves $a=0$ also
correspond to the results of calculation in Fig. \ref{fg6} based on Eqs. (%
\ref{ms330}) and (\ref{ms337}) - (\ref{ms340}). The figures suggest
convincingly, that the effect of finite $Q^2$ even for its rather low value,
can play some role in our integrals (\ref{app29}) - (\ref{app32}) only for
the higher effective masses ($x_0\simeq 0.5$), which are in our approach
evidently on the tail of the mass distribution $\mu _V(x_0)$. It means, that
the approximation (\ref{app0}) is for the purpose of our model fully
acceptable.



\newpage


\begin{thebibliography}{99}
\bibitem{jer}  {\it Proceedings of the XXIX International Conference on High
Energy Physics (ICHEP98), } UBC, Vancouver, B.C. Canada, July 23-29, 1998.

\bibitem{dis98}  {\it Proceedings of the 6th International Workshop on Deep
Inelastic Scattering and QCD - DIS98, }Brussels, Belgium, April 4-6, 1998.

\bibitem{ans}  M.~Anselmino, A.~Efremov, F.~Leader, Phys. Rep. {\bf 261}, 1
(1995).

\bibitem{hai}  Hai-Yang Cheng, Int. J. Mod. Phys. A {\bf 11}, 5109 (1996).

\bibitem{zav1}  P.~Zavada, Phys. Rev. {\bf D55}, 4290 (1997).

\bibitem{zav2}  P.~Zavada, Phys. Rev. {\bf D56}, 5834 (1997).

\bibitem{bo1}  Bo-Qiang Ma, J. Phys. G: Nucl. Part. Phys. {\bf 17}, L53
(1991).

\bibitem{bo2}  Bo-Qiang Ma and Qi-Ren Zhang, Z. Phys. {\bf C 58}, 479 (1993).

\bibitem{bo3}  Bo-Qiang Ma, Phys. Lett. {\bf B 375}, 320 (1996).

\bibitem{bo4}  Bo-Qiang Ma, {\it Proceedings of the 3rd Workshop on
Diquarks, Torino, Italy, October 28-30, 1996;} (World Sci.), edited by M.
Anselmino and E. Predazzi (in press).

\bibitem{rit}  Liang Zuo-tang and R. Rittel, Mod. Phys. Lett. {\bf A12}, 827
(1997).

\bibitem{jack}  J.D. Jackson, G.G. Ross, R.D. Roberts, Phys. Lett. {\bf B226}%
, 159 (1989).

\bibitem{bjo}  J.B.~Bjorken, Phys. Rev. {\bf 148}, 1467 (1966); {\it ibid.} 
{\bf D1}, 1376 (1970).

\bibitem{eja}  J.~Ellis and R.L.~Jaffe, Phys. Rev. {\bf D9}, 1444 (1974); 
{\it ibid.} {\bf D10}, 1669 (1974).

\bibitem{f2h1}  H1 Collaboration (T.~Ahmed {\it et al.}) Nucl. Phys. {\bf %
B439}, 471 (1995).

\bibitem{smc}  SMC Collaboration (B.~Adeva {\it et al.}) Phys. Lett. {\bf %
B412}, 414 (1997).

\bibitem{smcd}  SMC Collaboration (D.~Adams {\it et al.}) Phys. Rev. {\bf D56%
}, 5330 (1997).

\bibitem{ww}  S.~Wandzura and W.~Wilczek, Phys. Lett. {\bf B72}, 195 (1977).

\bibitem{msr}  A.D.~Martin, W.J.~Stirling, R.G.~Roberts, Phys. Rev. {\bf D50}%
, 6734 (1994).

\end{thebibliography}
\end{document}